\documentclass[conference]{IEEEtran}

\usepackage{booktabs}
\usepackage{color}
\usepackage{url}
\usepackage{tabularx}
\usepackage{ragged2e}
\usepackage{amsmath,amssymb,amsfonts}
\usepackage{siunitx}
\usepackage{multirow}
\usepackage{tikz}
\usepackage{pgfplots}
\usepackage[super]{nth}
\usepackage{array}
\usepackage{cite}
\usetikzlibrary{positioning}

\def\BibTeX{{\rm B\kern-.05em{\sc i\kern-.025em b}\kern-.08em
    T\kern-.1667em\lower.7ex\hbox{E}\kern-.125emX}}
\begin{document}

\tikzstyle{startstop} = [rectangle, rounded corners, minimum width=1cm, minimum height=1cm,text centered, text width=1.5cm,font = \scriptsize,draw=black, fill=red!10]
\tikzstyle{process} = [rectangle, minimum width=1cm, minimum height=1cm, text centered, text width=1.5cm, font = \scriptsize, draw=black, fill=gray!10]
\tikzstyle{decision} = [diamond, minimum width=3cm, minimum height=1cm, text centered, draw=black, fill=green!30]
\tikzstyle{arrow} = [thick,->,>=stealth]

\definecolor{KITDarkGreen}{HTML}{009682}
\definecolor{KITLightGreen}{HTML}{4664AA}
\definecolor{KITDarkGray}{HTML}{404040}

\title{Transformer-Based Rhythm Quantization of Performance MIDI Using Beat Annotations}

\author{
  \IEEEauthorblockN{Maximilian Wachter}
  \IEEEauthorblockA{\textit{Klangio GmbH}\\
    Karlsruhe, Germany}
  \and
  \IEEEauthorblockN{Sebastian Murgul}
  \IEEEauthorblockA{\textit{Klangio GmbH}\\
    Karlsruhe, Germany}
  \and
  \IEEEauthorblockN{Michael Heizmann}
  \IEEEauthorblockA{\textit{Institute of Industrial Information Technology} \\
    \textit{Karlsruhe Institute of Technology}\\
    Karlsruhe, Germany}
}

\maketitle

\begin{abstract}
  Rhythm transcription is a key subtask of notation-level Automatic Music Transcription (AMT). While deep learning models have been extensively used for detecting the metrical grid in audio and MIDI performances, beat-based rhythm quantization remains largely unexplored.
In this work, we introduce a novel deep learning approach for quantizing MIDI performances using a priori beat information. Our method leverages the transformer architecture to effectively process synchronized score and performance data for training a quantization model. Key components of our approach include dataset preparation, a beat-based pre-quantization method to align performance and score times within a unified framework, and a MIDI tokenizer tailored for this task.
We adapt a transformer model based on the T5 architecture to meet the specific requirements of rhythm quantization.
The model is evaluated using a set of score-level metrics designed for objective assessment of quantization performance. Through systematic evaluation, we optimize both data representation and model architecture. Additionally, we apply performance and score augmentations, such as transposition, note deletion, and performance-side time jitter, to enhance the model's robustness. Finally, a qualitative analysis compares our model's quantization performance against state-of-the-art probabilistic and deep-learning models on various example pieces.
Our model achieves an onset F1-score of $\SI{97.3}{\percent}$ and a note value accuracy of $\SI{83.3}{\percent}$ on the ASAP dataset. It generalizes well across time signatures, including those not seen during training, and produces readable score output. Fine-tuning on instrument-specific datasets further improves performance by capturing characteristic rhythmic and melodic patterns.
This work contributes a robust and flexible framework for beat-based MIDI quantization using transformer models.
\end{abstract}

\section{Introduction}
Automatic Music Transcription (AMT) is one task of specific interest in the field of Music Information Retrieval with the goal of retrieving a symbolic music representation from an input audio signal \cite{benetos_automatic_2019}.
In most research available the representation retrieved is in a MIDI format with unquantized note onset and offset times.
In order to retrieve a notation-level transcription that can be displayed as sheet music, the note timing has to be quantized to an underlying beat grid.
This beat grid can be derived from the result of a beat tracking approach or adopted directly in case a piece is performed to a metronome.
Since existing probabilistic and deep-learning-based quantization models infer beat information from transcribed rhythms, they are not capable of leveraging metronomic data, even if it is available.
Our model on the other hand utilizes this beat grid as a priori information.
Furthermore, while it would be possible to directly calculate the quantized note durations from the beat grid, the resulting durations would still reflect all human performance inaccuracies. Therefore, the result would only be barely readable and not practical in most applications.
\\\\\indent
In this work, we propose a new transformer-based approach that is able to accurately quantize performance MIDIs to scores based on a priori beat information. We show that this approach offers flexibility as well as control over the outputs by simply encoding time signature changes in the beat counter. By fine-tuning the provided model on instrument-specific datasets we are capable of further optimizing the results by modeling common rhythmic and melodic patterns that are used for each instrument.
\section{Related Work}
While beat detection has been extensively studied, the field of rhythm quantization remains comparatively underexplored. This section distinguishes between research efforts focused on beat tracking and those addressing rhythm quantization.
%
%
Beat tracking aims to detect the temporal positions of beats within a musical performance.
In 2011, Böck et al.\ introduced a frame-level beat classification method using bidirectional Recurrent Neural Networks (RNNs) and autocorrelation smoothing \cite{bock_enhanced_2011}.
This method was later extended with a Dynamic Bayesian Network (DBN) for modeling meter and bar structure \cite{bock_joint_2016}.
In 2019, Davies et al.\ improved this by replacing RNNs with a dilated Temporal Convolutional Network (TCN) \cite{davies_temporal_2019}.
Zhao et al.\ proposed Beat Transformer in 2022, incorporating attention mechanisms and demixed spectrograms \cite{zhao_beat_2022}.
Foscarin et al.\ introduced Beat This! in 2024, a transformer model robust to style and tempo changes, eliminating the need for DBN postprocessing \cite{foscarin_beatthis_2024}.
Most recently, in 2025, Murgul et al.\ reframed beat tracking in performance MIDI as a transformer-based sequence translation task \cite{murgul_beat_2025}.
\\\\\indent
%
Rhythm quantization involves aligning performed note onsets to a metrical grid to obtain a symbolic music representation. Early methods relied on rule-based and probabilistic techniques, while more recent approaches leverage machine learning models.
Cambouropoulos et al.\ proposed a system for joint beat detection and rhythm quantization in 2000 \cite{cambouropoulos_midi_2000}. Their approach clustered inter-onset intervals for beat detection, followed by assigning note onsets to the closest points on a metrical grid and assigning note values based on inter-onset intervals. Cemgil et al.\ introduced a quantization framework in the same year, utilizing Bayesian probabilistic modeling, incorporating a performance model to formalize simple quantization strategies alongside a prior model to account for rhythmic complexity \cite{cemgil_rhythm_2000}.
In 2002, Takeda et al.\ proposed the first method utilizing Hidden Markov Models (HMMs) for the rhythm transcription task \cite{takeda_hidden_2002}. They employed the Viterbi algorithm to estimate note values by combining a stochastic model of timing deviations and a grammatical model of plausible note sequences. Hamanaka et al.\ proposed a method in 2003 for estimating intended onset times from fixed-tempo jam sessions by training an HMM with human performance data using the Baum-Welch algorithm \cite{hamanaka_learning-based_2003}.
Temperley's 2007 book \lq{}Music and Probability\rq{} extended Bayesian probabilistic approaches to infer complete metrical grids rather than score positions relative to a bar \cite{temperley_music_2007}.
Cogliati et al.\ presented an HMM-based system in 2016 for joint estimation of meter, harmony, and stream separation, combined with a distance-based quantization algorithm \cite{cogliati_transcribing_2016}.
Foscarin et al.\ introduced a parse-based system in 2019 employing weighted context-free grammars (WCFGs) for joint rhythm quantization and music score production \cite{montiel_parse-based_2019}.
Shibata et al.\ proposed a piano transcription system in 2021 that incorporated HMMs and Markov Random Fields (MRFs) for rhythm quantization, leveraging non-local musical statistics to infer global parameters \cite{shibata_non-local_2021}.
Liu et al.\ proposed a Convolutional-Recurrent Neural Network (CRNN)-based system in 2022 for MIDI-to-score conversion, incorporating onset-based beat detection and rhythm quantization \cite{liu_performance_2022}.
Kim et al.\ developed a transformer and Convolutional Neural Network (CNN)-based guitar transcription model in 2023 that produced note-level transcriptions from spectrograms using beat information \cite{kim_note-level_2022}.
Beyer et al.\ introduced a performance MIDI-to-score conversion approach in 2024 based on the Roformer architecture. Their encoder-decoder model directly generated MusicXML tokens while implicitly performing beat estimation and rhythm quantization on MIDI token sequences \cite{beyer_end_2024}.
\\\\\indent
Although recent transformer-based models have advanced MIDI-to-score conversion, most rely on end-to-end architectures with implicit beat estimation. This approach prevents the use of external beat information, such as metronome data or manually annotated beats. Incorporating explicit beat inputs into quantization systems improves both flexibility and interpretability, filling an important gap in current research.
\section{Methodology}
\subsection{Task Definition}
Our model aims to convert an unquantized performance MIDI sequence $X_n$, represented in the time domain, into a score-like sequence $Y_n$ with musical timing information, guided by beat and downbeat annotations $X_\text{b}$ and $X_\text{db}$.
Unlike most state-of-the-art quantization models, such as \cite{beyer_end_2024}, which operate solely on time-domain input, our approach explicitly incorporates a priori beat information, including beat estimations or metronomic data. In particular, when a performance is aligned to a metronome, this information can remove ambiguity about the underlying metrical grid, leading to more accurate quantization.
The model assumes a one-to-one correspondence between performance and score notes, ensuring that no additional notes are added or removed during quantization. This criterion further ensures correspondence between measures in scores and performances, which is necessary due to the at times poor alignment between them.

The input sequence $X_n$ is defined as
\begin{equation}
  X_n = \{(p_i, o_i, d_i)\}^{N_\text{perf}}_{i=1}
\end{equation}
with the individual notes in the sequence being represented by the MIDI pitch $p_i$, onset $o_i$, and duration $d_i$ in seconds.
Notes in the target sequence $Y_n$ are represented using musical onsets $\textit{mo}_i$ and musical note values $\textit{mnv}_i$ described by
\begin{equation}
  Y_n = \{(p_i, \textit{mo}_i, \textit{mnv}_i, n_\text{measure})\}^{N_\text{perf}}_{i=1} \quad.
\end{equation}
To limit the range of possible values, musical onsets hereby denote a note's position within its respective measure instead of the absolute position within the entire piece. Thus, a note's position within a score is defined by its measure number $n_\text{measure}$ and its musical onset time.

Beat annotations $X_\text{b}$ and downbeat annotations $X_\text{db}$ are given in the form of
\begin{equation}
  X_\text{b} = \{(t_j)\}^{N_\text{beat}}_{j=1}
\end{equation}
and
\begin{equation}
  X_\text{db} = \{(t_k)\}^{N_\text{downbeat}}_{k=1}
\end{equation}
where $X_\text{db} \subseteq X_\text{b}$.

While it is logical to train the model using ground truth beat data, it is reasonable to assume that, once a model has gained a deep understanding of rhythmic structure, it is able to produce a meaningful quantization even with faulty beat data.
To obtain a tokenizable representation of $X_n$, the MIDI note time information is fused with $X_\text{b}$ and $X_\text{db}$. Therefore, we interpolate $X_\text{b}$ to twelve equidistant sub-beats (referred to as ticks $\textit{tick}_l$) per beat, obtaining a \nth{32}-note triplet grid $X_\textit{ticks}$ given as
\begin{equation}
  X_\textit{ticks} = \{(\textit{tick}_l)\}^{12\cdot N_\text{beat}}_{l=1}.
\end{equation}
This resolution is chosen as it allows representation of straight and triplet-based note values of a \nth{16}-note triplet and above.
The continuous onset and duration times are then quantized to \nth{32}-note triplets using the Euclidean distance given by
\begin{equation}
  \underset{l}{\arg\min} \, \left( | \textit{tick}_l - o_i | \right) \quad.
\end{equation}

To enable independent training on individual measures, onset values are reset to zero at the beginning of each measure. While this step already converts performance note times into musical note values, the results are not suitable for human-readable scores due to timing inconsistencies in expressive performances, which often produce irregular or complex note durations.
The score sequences are based on the MusicXML format \cite{makemusic_musicxml_2025}, where onset and note values are expressed in quarter note units. To align with this representation, onsets and durations are scaled by a factor of $12$. With both $X_n$ and $Y_n$ expressed in musical note values, input and target sequences can be encoded using a unified tokenization scheme.

\subsection{Tokenization Scheme}
\label{sec:token}
The tokenization scheme is designed to efficiently encode the note sequences derived from both performance and score data. It is loosely based on the approach introduced in \cite{hawthorne_sequence_2021}. In our model, each note in the input sequence $X_n$ and the target sequence $Y_n$ is represented using three distinct tokens: one each for pitch, onset, and note value.
Measure numbers are not encoded explicitly. Instead, the start of a new measure is indicated by a dedicated \lq{}new measure\rq{} token, eliminating the need for a wide range of measure number tokens. For input sequences, a new measure token is inserted when a note's onset exceeds the downbeat of the following measure. In the target sequences, new measure tokens are inferred directly from the score.
The vocabulary consists of $88$ pitch tokens (covering the full range of piano MIDI pitches), 48 onset tokens (corresponding to \nth{32}-note triplet subdivisions in a 4/4 measure), and $48$ note value tokens. As a result, the current model supports quantization of measures up to the length of a whole note, with note values capped at that duration. The full set of token types and their ranges is detailed in Table~\ref{tab:token}.

\begin{table}
  \centering
  \caption{Summary of token types used in the model's vocabulary, including pitch, onset, note value, and structural indicators, with corresponding value ranges and total counts.}
  \begin{tabularx}{0.49\textwidth}{X|X|X}
    \hline
    \textbf{Token Type} & \textbf{No. of Values} & \textbf{Value Range} \\
    \hline
    Pitch               & $88$                   & $[21, 108]$          \\
    Onset               & $48$                   & $[0, 47]$            \\
    Note value          & $48$                   & $[1, 48]$            \\
    New measure         & $1$                    & $\{0\}$              \\
    \hline
  \end{tabularx}
  \label{tab:token}
\end{table}

\subsection{Network Architecture}
Our model architecture is based on the T5 transformer by Raffel et al.\ \cite{raffel_exploring_2020}, but we omit pre-training since we use a custom token vocabulary described in Section~\ref{sec:token}. The configuration is significantly smaller than \textit{t5-small}.
Specifically, the model uses a key/value dimensionality of $d_\text{kv} = 64$, a feed-forward layer size of $d_\text{ff} = 1024$, and a vocabulary size of 187, which includes the tokens from Section~\ref{sec:token} along with end-of-sequence (EOS) and padding (PAD) tokens. During optimization, we reduced the number of layers from six to two, the number of attention heads from eight to four, and the embedding size from $512$ to $128$. This configuration yielded the best empirical results.
The improved performance of the smaller model can be attributed to the structured nature of onset and note value data, where closely related values frequently co-occur, and to the compact vocabulary size, which renders larger embedding spaces unnecessary. As a result, our model is more computationally efficient than other state-of-the-art approaches. Furthermore, by processing short segments of $M$ measures sequentially, computational cost grows linearly with input length rather than exponentially, enabling scalable and parallelizable inference.

\subsection{Training and Inference}
The T5 model is trained using cross-entropy loss and the Adafactor optimizer \cite{shazeer_adafactor_2018}. Although prior work such as \cite{hawthorne_sequence_2021} recommends a fixed learning rate of $0.001$, we adopt an adaptive learning rate, which led to faster and more stable convergence in our experiments.
Each input sequence consists of $M$ measures, where $M$ is a tunable parameter (see Section~\ref{sec:opti}). The sequences are non-overlapping, as empirical results showed improved performance with shorter inputs, indicating limited long-range context dependency. Notes are tokenized into a single one-dimensional sequence, with each note represented by an ordered triplet of pitch, onset, and note value tokens. This requires the model to learn the correct token structure to generate processable outputs.
Training was performed for up to $100$ epochs with a batch size of eight. An early stopping criterion was applied, terminating training if validation loss did not improve for $20$ consecutive epochs. In practice, convergence typically occurred before epoch $60$. We used a dropout rate of $0.1$ for regularization during training. During inference, we apply beam search decoding with a beam width of five.

\section{Experiments}
\subsection{Datasets}
Effective rhythm transcription requires learning the relationship between expressive timing variations in human performances and the corresponding notated musical timing. Consequently, a suitable dataset must include ground truth scores, human-performed MIDI recordings, and precise beat and downbeat annotations.
These criteria exclude the widely used A-MAPS dataset \cite{ycart_amaps_2018}, as its performances are generated from tempo-modified quantized MIDI files and do not reflect the full spectrum of human timing deviations \cite{foscarin_asap_2020}. This leaves only a small selection of datasets.

Firstly, we use the ASAP dataset \cite{foscarin_asap_2020}, which, to the best of our knowledge, most effectively meets these requirements. It contains $1,067$ performance MIDI files spanning $236$ classical piano pieces, with many pieces having multiple performances. Each performance is paired with a MusicXML score and includes annotations for beats, downbeats, time signatures, and musical keys.

\subsection{Dataset Preparation}
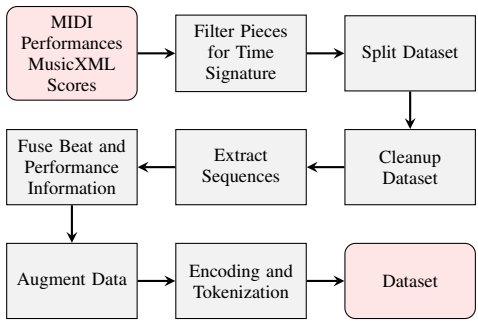
\begin{figure}[h]
  \centering
  \begin{tikzpicture}[node distance=0.5cm and 0.5cm]

    \node (start) [startstop] {MIDI Performances MusicXML Scores};
    \node (filter) [process, right=of start] {Filter Pieces for Time Signature};
    \node (split) [process, right=of filter] {Split Dataset};
    \node (cleanup) [process, below=of split] {Cleanup Dataset};
    \node (sequences) [process, left=of cleanup] {Extract Sequences};
    \node (info) [process, left=of sequences] {Fuse Beat and Performance Information};
    \node (augment) [process, below=of info] {Augment Data};
    \node (token) [process, right=of augment] {Encoding and Tokenization};
    \node (stop) [startstop, right=of token] {Dataset};

    \draw [arrow] (start) -- (filter);
    \draw [arrow] (filter) -- (split);
    \draw [arrow] (split) -- (cleanup);
    \draw [arrow] (cleanup) -- (sequences);
    \draw [arrow] (sequences) -- (info);
    \draw [arrow] (info) -- (augment);
    \draw [arrow] (augment) -- (token);
    \draw [arrow] (token) -- (stop);

  \end{tikzpicture}
  \caption{Overview of the data preprocessing steps, from raw MIDI and MusicXML files to finalized token sequences. Including filtering, alignment, augmentation, and tokenization stages for training-ready inputs.}
  \label{fig:data_prep}
\end{figure}

Figure~\ref{fig:data_prep} illustrates the data preparation pipeline for our model. We train on measures in the time signatures 4/4, 3/4, and 2/4, which are separated into distinct datasets to facilitate comparison. Other time signatures are omitted. Aside from this separation, the model is time signature-agnostic, as the relevant time signature is provided as a priori information and applied during preprocessing. This results in $85$ pieces with 4/4 measures, $53$ with 3/4, and $50$ with 2/4.

We randomly select $\SI{10}{\percent}$ of the pieces for the test and validation sets, using the remaining $\SI{90}{\percent}$ for training. Since ties in MusicXML are represented as two distinct notes, we resolve them by matching each tie end to its corresponding tie start, merging their durations and removing the redundant note.
Training examples are extracted as sequences of $M$ measures. To simplify input, hand separation is removed so that each sequence consists of a single stream of individual notes. Measures whose actual duration does not match the annotated time signature are excluded.
As our model assumes a one-to-one correspondence between performance and score notes, we include only those measures where the number of notes in both sequences aligns. To increase the number of eligible measures, we shift the search interval for note matching \SI{50}{\milli\second} earlier than the annotated downbeats. This accounts for notes played slightly ahead of the measure boundary, which typically belong to the following measure. The \SI{50}{\milli\second} threshold was chosen empirically to maximize training coverage.
After filtering and alignment, approximately $40,000$ measures remain available for training.

We similarly adapt the model to guitar data using the Leduc dataset \cite{edwards_francois_2024}, which contains 239 jazz guitar performances with high-quality transcriptions by François Leduc in GuitarPro\footnote{\url{https://www.guitar-pro.com}} format. These scores are converted to MIDI and aligned with the original audio using the method described by Riley et al.\ \cite{riley_high_2024} which we extended to also align beats and downbeats. We then apply the same preprocessing pipeline used for the piano data, treating the aligned MIDI transcriptions as performance input. 

\subsection{Metrics}
We use two sets of metrics for evaluation: one for model optimization and one for comparison with other methods. Since the model is trained under the assumption of a one-to-one correspondence between performance and score notes, we ensure that the length of the generated sequence matches the unquantized input at inference time.

For optimization, we use onset F1-score, note value accuracy, and note value mean squared error (MSE). A true positive in onset F1 is defined as a note with exactly matching pitch and onset in both the quantized output and the ground truth. This metric relies on ground truth beat annotations, as any deviation would misalign onsets; therefore, it cannot be used to compare models that do not leverage beat information and are therefore not alignable.
Note value metrics are computed only for notes with correctly predicted onsets, as correct note values at incorrect positions are not musically meaningful. This introduces a bias: errors in onset prediction often imply errors in note duration. Note value MSE, expressed in quarter note units, is averaged over all test examples. As note value accuracy is strongly tied to onset precision, it serves as the primary metric for model tuning.

For broader evaluation and comparison, we adopt the edit-distance-based metrics introduced in \cite{nakamura_towards_2018}, commonly referred to as MUSTER scores \cite{beyer_end_2024}. The onset-time error rate $\epsilon_{onset}$ is calculated based on the number of shift and scaling operations needed to align the estimated score with the ground truth \cite{nakamura_rhythm_2017}. The offset-time error rate $\epsilon_{offset}$ reflects mismatches in note offsets relative to their onsets.
To account for additional, missing, or incorrect notes, the output and reference sequences are first aligned, also yielding a pitch error rate $\epsilon_p$, an extra note rate $\epsilon_{extra}$, and a missing note rate $\epsilon_{miss}$.
However, since our model assumes a one-to-one correspondence between input and output notes, these three metrics are not applicable and are therefore excluded from evaluation.

\subsection{Model Optimization}
\label{sec:opti}
Our optimization strategy begins with training a baseline model using only 4/4 measures from the ASAP dataset, focusing on tuning model parameters and evaluating data augmentation methods.
Once optimized, the model is extended to handle additional time signatures and, finally, adapted to guitar performances from the Leduc dataset.

\subsubsection{Optimizing Training Sequences}
We begin by optimizing the number of measures per input sequence ($M$) and the ordering of notes within each sequence. As shown in Table~\ref{tab:measures}, using two-measure sequences yields slightly better performance than other configurations.
Apparently, the two-measure version provides sequences that are short enough to learn meaningful representations in the intermediate layers but still provides enough contextual information between the measures.

Further improvements are achieved by synchronizing the note order between input and target sequences, leveraging the one-to-one correspondence assumption. Specifically, we reorder the target sequence to match the onset-sorted order of the input sequence, ensuring a more consistent mapping between corresponding notes. As Table~\ref{tab:measures} indicates, this synchronization results in a notable increase of approximately $\SI{2}{\percent}$ in onset F1-score compared to unsynchronized sequences.

\begin{table}[t]
  \centering
  \caption{Evaluation results on the ASAP dataset for models trained with one to four measures per input sequence, with and without synchronized note ordering. Metrics include Onset F1-score, Note Value Accuracy (NV Acc.), and Note Value Mean Squared Error (NV MSE).}
  \begin{tabularx}{\columnwidth}{X|X|>{\raggedleft\arraybackslash}X>{\raggedleft\arraybackslash}X>{\raggedleft\arraybackslash}X}
    \hline
    \textbf{Number of Measures} & \textbf{Syn\-chro\-nized} & \textbf{Onset F1}              & \textbf{NV Acc.}               & \textbf{NV MSE} \\ \hline 
    \textbf{One}                & No                        & \SI{93.4 }{\percent}           & \SI{80.9  }{\percent}          & 0.25            \\
    \textbf{One}                & Yes                       & \SI{95.8   }{\percent}         & \SI{81.8  }{\percent}          & \textbf{0.21}   \\
    \textbf{Two}                & No                        & \SI{94.0  }{\percent}          & \SI{82.5   }{\percent}         & 0.23            \\
    \textbf{Two}                & Yes                       & \textbf{\SI{96.0 }{\percent} } & \textbf{\SI{82.6 }{\percent} } & \textbf{0.21}   \\
    \textbf{Three}              & Yes                       & \SI{95.9    }{\percent}        & \SI{80.4     }{\percent}       & 0.27            \\
    \textbf{Four}               & Yes                       & \SI{94.8  }{\percent}          & \SI{81.2    }{\percent}        & 0.22            \\
    \hline
  \end{tabularx}
  \label{tab:measures}
\end{table}

\subsubsection{Data Augmentation}
To further enhance model performance, we apply three data augmentation strategies during training:
\begin{itemize}
  \item \textbf{Transposition:} input and target sequences are randomly transposed by a fixed number of semitones. The transpose value is drawn from a uniform distribution so that the available MIDI pitch range for piano is not exceeded.
  \item \textbf{Deletion:} During training $\SI{20}{\percent}$ of notes are randomly selected and deleted from the input and label sequences with a $\SI{50}{\percent}$ probability.
  \item \textbf{Note Value Noise:} Since the deviation between performance and score is quite high for note durations, we add a noise term to performance note durations, following a normal distribution with a standard deviation of $\SI{5}{\percent}$ of the note duration.
\end{itemize}

\begin{figure}[h]
  \centering
  \begin{tikzpicture}
    \begin{axis}[
        ybar,
        bar width=0.2cm,
        width=\columnwidth,
        height=0.35\textwidth,
        ymajorgrids = true,
        symbolic x coords={-, T, D, N, T/D, T/N, D/N, All},
        xtick=data,
        nodes near coords={},
        ymin=75,
        ymax=105,
        yticklabel={\pgfmathparse{\tick}\pgfmathprintnumber{\pgfmathresult}\%},
        enlarge x limits=0.15,
        enlarge y limits=0,
        x tick label style={text width=1em, inner sep=2pt, align=center},
        legend style={at={(0.95,0.95)}, anchor=north east,legend columns=-1} 
      ]
      \addplot[fill=KITDarkGreen] coordinates {(-, 96.0) (T, 97.0) (D, 96.3) (N, 96.2) (T/D, 96.2) (T/N, 96.8) (D/N, 96.7) (All, 96.5)};
      \addplot[fill=KITLightGreen] coordinates {(-, 82.6) (T, 83.5) (D, 82.9) (N, 81.6) (T/D, 83.9) (T/N, 84.2) (D/N, 82.9) (All, 84.0)};
      \legend{$Onset F1$, $NV Acc$}
    \end{axis}
  \end{tikzpicture}
  \caption{Comparison of Onset F1-score and Note Value Accuracy in percent for various combinations of augmentation methods: Transposition (T), Deletion (D), and Note Value Noise (N). Combining methods yields the best overall performance.}
  \label{fig:aug_eval}
\end{figure}
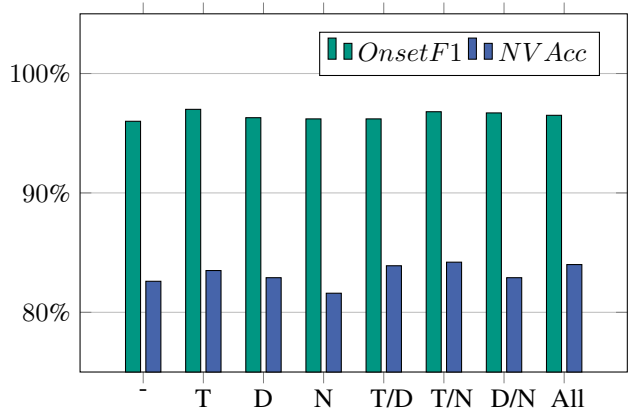

The effects of these augmentations on onset F1-score and note value accuracy are shown in Figure~\ref{fig:aug_eval}.

Notably, all augmentation methods combined improve the onset F1-score. However, applying note value noise alone reduces note value accuracy. When combined with other augmentations, especially transposition, it yields substantial improvements. This may be due to delayed early stopping, allowing the model more time to generalize from the augmented data.
Interestingly, transposition improves quantization performance even though it does not modify rhythm. This suggests it strengthens the model's understanding of pitch-related structure and exposes it to underrepresented pitch ranges.
Based on these findings, subsequent models are trained using a combination of transposition and note value noise, as this yields the best balance between onset and note value accuracy.

\subsubsection{Extending to Different Time Signatures}
\label{sec:time_signatures}
Since our model implicitly handles time signatures, no explicit time signature token is required. Instead, support for different time signatures is achieved through the preprocessing procedure described in Section \ref{sec:token}, where note timing is given relative to the measure start but not tied to a specific time signature. As a result, the model can generalize to time signatures it was not explicitly trained on.
Table~\ref{tab:time signatures} presents evaluation results for models trained on various combinations of time signatures. Each model is tested on a separate set containing a single time signature to assess how including data from other meters influences performance.
\begin{table}[t]
  \centering
  \caption{Quantization results on the ASAP dataset for models trained on various combinations of 2/4, 3/4, and 4/4 time signatures and tested on each time signature individually. Including multiple time signatures generally improves performance and generalization.}
  \begin{tabularx}{\columnwidth}{X|m{1.8cm}|>{\raggedleft\arraybackslash}X>{\raggedleft\arraybackslash}X>{\raggedleft\arraybackslash}X}
    \hline
    \textbf{Tested On}            & \multicolumn{1}{>{\centering\arraybackslash}m{1.8cm}|}{\textbf{Trained On}} & \textbf{Onset F1}            & \textbf{NV Acc.}             & \textbf{NV MSE} \\ \hline 
    \multirow{3}{*}{\textbf{4/4}} & 4/4                                                                         & \SI{ 96.8 }{\percent}        & \SI{84.2 }{\percent}         & 0.19            \\
                                  & 3/4, 4/4                                                                    & \textbf{\SI{96.9}{\percent}} & \textbf{\SI{85.2}{\percent}} & \textbf{0.17}   \\
                                  & 2/4, 3/4, 4/4                                                               & \SI{96.8   }{\percent}       & \SI{ 84.0      }{\percent}   & 0.20            \\
    \hline
    \multirow{4}{*}{\textbf{3/4}} & 3/4                                                                         & \SI{97.4   }{\percent}       & \SI{ 76.8   }{\percent}      & 0.47            \\
                                  & 4/4                                                                         & \SI{ 95.9   }{\percent}      & \SI{68.5   }{\percent}       & 0.54            \\
                                  & 3/4, 4/4                                                                    & \SI{ 97.9    }{\percent}     & \textbf{\SI{80.4}{\percent}} & 0.41            \\
                                  & 2/4, 3/4, 4/4                                                               & \textbf{\SI{98.1}{\percent}} & \SI{79.0    }{\percent}      & \textbf{0.40}   \\
    \hline
    \multirow{3}{*}{\textbf{2/4}} & 2/4                                                                         & \SI{96.0  }{\percent}        & \SI{ 80.8 }{\percent}        & 0.12            \\
                                  & 4/4                                                                         & \SI{ 96.7   }{\percent}      & \SI{84.2   }{\percent}       & 0.13            \\
                                  & 3/4, 4/4                                                                    & \SI{97.0  }{\percent}        & \SI{84.5    }{\percent}      & 0.13            \\
                                  & 2/4, 3/4, 4/4                                                               & \textbf{\SI{98.0}{\percent}} & \textbf{\SI{86.9}{\percent}} & \textbf{0.10}   \\
    \hline
  \end{tabularx}
  \label{tab:time signatures}
\end{table} 

Incorporating multiple time signatures, and therefore a more diverse dataset, generally improves quantization performance. This effect is likely due to increased data variety, which enhances the model's ability to generalize. Rhythmic patterns learned from one time signature often transfer successfully to others, as shown by the strong performance of the 4/4-trained model on both 2/4 and 3/4 sequences. These findings support the use of a single model across all time signatures rather than training separate models for each case.
An exception, as shown in Table~\ref{tab:time signatures}, is a slight decrease in note value accuracy for 4/4 when 2/4 data is included in training. This may be explained by the reduced occurrence of longer note values in shorter measures.
Nonetheless, the overall advantages of a combined model, including improved generalization and simplified deployment, make it the preferred approach.

\subsubsection{Training on Guitar Data}

Finally, we extend the model to guitar data using the Leduc dataset \cite{edwards_francois_2024}. To evaluate instrument-specific generalization, we train three versions of the model: one using only guitar data, one using only piano data, and one trained on a combined dataset.
Each model is tested on both guitar and piano data to assess whether rhythmic performance characteristics differ by instrument and whether cross-instrument generalization is feasible. All models are trained on measures in 2/4, 3/4, and 4/4 time signatures, consistent with the findings in Section~\ref{sec:time_signatures}.
The evaluation results are presented in Figure~\ref{tab:instr_eval}.

\begin{table}[h]
  \centering
  \caption{Onset F1-score and Note Value Accuracy for models trained on guitar (Leduc), piano (ASAP), and combined datasets, evaluated on both test sets. Instrument-specific models outperform cross-domain models, especially in note value accuracy.}
  \begin{tabularx}{0.49\textwidth}{l|l|>{\centering\arraybackslash}X|>{\centering\arraybackslash}X}
    \hline
    \textbf{Tested on}     & \textbf{Trained on} & \textbf{Onset F1}    & \textbf{Note value acc.} \\
    \hline
    \multirow{3}{*}{Leduc} & Leduc               & \SI{92.1}{\percent}  & \SI{90.2   }{\percent}   \\
                           & ASAP                & \SI{87.2}{\percent}  & \SI{71.3   }{\percent}   \\
                           & Leduc + ASAP        & \SI{90.3 }{\percent} & \SI{86.9   }{\percent}   \\
    \hline
    \multirow{3}{*}{ASAP}  & Leduc               & \SI{90.5 }{\percent} & \SI{69.4  }{\percent}    \\
                           & ASAP                & \SI{97.3 }{\percent} & \SI{83.3  }{\percent}    \\
                           & Leduc + ASAP        & \SI{97.2 }{\percent} & \SI{81.1}{\percent}      \\
    \hline
  \end{tabularx}
  \label{tab:instr_eval}
\end{table}

Although our model operates solely on symbolic data, the results clearly indicate that models trained exclusively on data from a specific instrument outperform those trained on other instruments.
While the combined model reliably yields improvements compared to the evaluation results for models trained without the evaluated instrument, it still underperforms compared to the evaluation results of the models trained exclusively on the evaluated instrument.
Differences in the quantization performance become especially apparent for note value accuracy, which corresponds with the assumption that the characteristics of rhythmic interpretation of note durations vary from instrument to instrument.
Consequently, training separate models for each instrument, when feasible, appears to be more effective than relying on a single, universal model.

\subsection{Comparative Experiments}

\begin{table}[b]
  \centering
  \caption{Comparison of onset-time ($\epsilon_\textit{onset}$) and offset-time ($\epsilon_\textit{offset}$) error rates using the MUSTER metric. The proposed model outperforms all baselines in onset quantization and ranks second in offset accuracy.}
  \begin{tabularx}{0.49\textwidth}{p{4cm}|>{\raggedleft\arraybackslash}X|>{\raggedleft\arraybackslash}X}
    \hline
    \textbf{Method}                                             & \multicolumn{1}{>{\centering\arraybackslash}X|}{\textbf{$\epsilon_\textit{onset}$}} & \multicolumn{1}{>{\centering\arraybackslash}X}{\textbf{$\epsilon_\textit{offset}$}} \\
    \hline 
    Neural Beat Tracking \cite{liu_performance_2022}            & 68.28                                                                             & 54.11                                                                             \\
    End-to-End PM2S \cite{beyer_end_2024}                       & 15.55                                                                             & \textbf{23.84}                                                                    \\
    HMMs + Heuristics (J-Pop) \cite{shibata_non-local_2021}     & 25.02                                                                             & 29.21                                                                             \\
    HMMs + Heuristics (classical) \cite{shibata_non-local_2021} & 22.58                                                                             & 29.84                                                                             \\
    MuseScore \cite{musescore}                                  & 47.90                                                                             & 49.44                                                                             \\
    Finale \cite{makemusic_finale_1988}                         & 31.85                                                                             & 45.34                                                                             \\
    \hline
    \textbf{Our Model}                                          & \textbf{12.30}                                                                    & 28.30                                                                             \\
    \hline
  \end{tabularx}
  \label{tab:comparison}
\end{table}

For comparative evaluation, we train our model using the ASAP splits defined in the ACPAS dataset \cite{liu2021acpas}, which combines predefined train/test splits from A-MAPS \cite{ycart_amaps_2018} and ASAP \cite{foscarin_asap_2020}. This setup ensures direct comparability with prior work, including \cite{liu_performance_2022} and \cite{beyer_end_2024}.
We use the MUSTER score \cite{nakamura_rhythm_2017} as our primary comparison metric by using the publicly available implementation found on Github\footnote{\url{https://github.com/amtevaluation/amtevaluation.github.io}}. Table \ref{tab:comparison} shows a comparison of our evaluation results with the results achieved by the deep learning-based model by Beyer et al.\ \cite{beyer_end_2024} as well as other models referenced in the same publication. These include the commercial products MuseScore \cite{musescore} and Finale \cite{makemusic_finale_1988} as well as state-of-the-art probabilistic and deep learning-based approaches \cite{shibata_non-local_2021, liu_performance_2022}.

In order to compute the metric, we quantize entire performances by segmenting the performances in the test set into sequences of two measures, quantizing the sequences, and concatenating them back to a quantized MusicXML score.
The comparative experiments pose an additional challenge, since the test set partly consists of time signatures that our model was not trained on, like 6/8 or even 12/16. We transcribe these by adapting the preprocessing for these time signatures to the beat count used in the annotations. For instance, since 6/8 measures are counted in two, we interpolate to 18 ticks per beat in this case.
To produce valid and readable scores, we apply post-processing steps including chord merging, tie reconstruction for notes crossing measure boundaries, and a simple voice separation algorithm to avoid overlapping notes within a voice.

Our model achieves the best performance in terms of $\epsilon_{onset}$ and is second only to \cite{beyer_end_2024} in $\epsilon_{offset}$. Notably, our model is not trained to recognize \nth{32} notes or irregular time signatures, which are present in the test set.
Despite this, the results demonstrate that leveraging beat annotations enables our model to match or surpass state-of-the-art quantization approaches.

\section{Conclusion}
In this work, we presented the first transformer-based model for beat-based rhythm quantization of MIDI performances. To enable this, we introduced a simple yet effective preprocessing method that fuses beat and performance information into a unified tokenized representation for both input and target sequences.
We adapted this preprocessing method to different time signatures and proved that the resulting model is capable of quantizing unseen time signatures once adapted to the preprocessing framework. We defined a confusion-based metric for evaluating beat-based quantization derived from musical onset and note value and optimized our model using different sequence structures and augmentations. While initially training on piano performances from the ASAP dataset, we adapted the model to the domain of guitar data using the Leduc dataset and showed that instrument-specific rhythm quantization models show better performance which is likely due to the differences in rhythmic interpretation between instruments.

Future work may focus on expanding the model's capabilities by incorporating a broader range of time signatures and note values, extending the dataset, and integrating additional musical context such as voice separation by including voice tokens to individual notes or explicit time signature tokens.

\bibliographystyle{IEEEtran}

\bibliography{root}

\end{document}